\documentclass[12pt]{article}
\newcommand{\Tr}{\mbox{Tr\,}}
\newcommand{\beq}{\begin{equation}}
\newcommand{\eeq}[1]{\label{#1}\end{equation}}
\newcommand{\bea}{\begin{eqnarray}}
\newcommand{\eea}[1]{\label{#1}\end{eqnarray}}

\newcommand{\newc}{\newcommand}
\newc{\gsim}{\lower.7ex\hbox{$\;\stackrel{\textstyle>}{\sim}\;$}}
\newc{\lsim}{\lower.7ex\hbox{$\;\stackrel{\textstyle<}{\sim}\;$}}
\newc{\gev}{\,{\rm GeV}}
\newc{\mev}{\,{\rm MeV}}
\newc{\ev}{\,{\rm eV}}
\newc{\kev}{\,{\rm keV}}
\newc{\tev}{\,{\rm TeV}}
\newc{\mz}{m_Z}
\newc{\mpl}{M_{pl}}
\renewcommand{\phi}{\varphi}
\newc\order{{\cal O}}
\newc\CO{\order}
\newc\CL{{\cal L}}
\newc{\eps}{\epsilon}
\newc{\re}{\mbox{Re}\,}
\newc{\im}{\mbox{Im}\,}
\newc{\invpb}{\,\mbox{pb}^{-1}}
\newc{\invfb}{\,\mbox{fb}^{-1}}

\begin{document}

\begin{titlepage}
\begin{flushright}
{LBL-46886 UCB-PTH-00/32 \\
NYU-TH/00/09/01 hep-ph/0012148\\
December 2000\\
}
\end{flushright}
\vskip 2cm
\begin{center}
{\large\bf Holography and Phenomenology}
\vskip 1cm
{\normalsize
Nima Arkani-Hamed$^a$, Massimo Porrati$^b$ and Lisa Randall$^c$}\\
\vskip 0.5cm
{\it a) Department of Physics, University of California\\
Berkeley, CA~~94720, USA\\
and \\
Theory Group, Lawrence Berkeley National Laboratory\\
Berkeley, CA~~94720, USA\vskip .1in
b) Department of Physics, New York University\\ 4 Washington Pl\\
New York, NY~~10003, USA\vskip .1in
c) Center for Theoretical Physics, MIT\\ Cambridge, MA 02139, USA\\
}
\end{center}
\vskip .5cm
\begin{abstract}
We examine various aspects of the conjectured duality between
warped AdS$_5$ geometries with boundary branes
 and strongly coupled (broken) conformal field theories
coupled to dynamical gravity. We also examine compactifications with 5-d
gauge fields, in which case the holographic dual is a broken CFT weakly
coupled to dynamical gauge fields in addition to gravity.
The holographic picture is used to
clarify a number of important phenomenological issues in these and related
models, including the questions of black hole production, radius
stabilization, early universe cosmology, and gauge coupling unification.

\end{abstract}
\end{titlepage}
\setcounter{footnote}{0}
\setcounter{page}{1}
\setcounter{section}{0}
\setcounter{subsection}{0}
\setcounter{subsubsection}{0}

\section{Introduction}
The lesson from the AdS/CFT correspondence~\cite{Juan,GKP,Ed} is
that a gravity theory on  $AdS_5$ is equivalent to a strongly
coupled 4D CFT. However, when all the symmetries are fully
realized, the physical equivalence between the theories is
slightly obscure. Normally, we think of two theories as being
equivalent if they make identical predictions for all physical
processes, such as the spectrum or S-matrix elements. However in
standard AdS/CFT, the theories on both sides are a little removed
from our usual intuition; on the 4D side, CFT's do not have
S-matrices or a ``spectrum'', and the same is true on the AdS
side. The equivalence takes on a slightly more abstract form; for
every CFT operator, ${\cal O}$, there is a corresponding bulk
field $\phi$. Given any boundary condition for $\phi$,
$\phi_0(x)$, at the 4D boundary of $AdS$, there is a unique
solution of the gravity (or more generally the full string)
equations of motion in the bulk; let $\Gamma[\phi_0(x)]$ represent
the SUGRA action (or more generally full string effective action)
of this solution. The AdS/CFT correspondence tells us that 
\beq
\langle \mbox{exp}(-\int d^4 x \phi_0  {\cal O}) \rangle_{CFT} =
\mbox{exp}(-\Gamma[\phi_0]). \eeq{n1} 
This correspondence does
contain all the physical information we can extract out of the
theories on both sides; however both sides are a little
unfamiliar.

In this respect, the addition of the Randall-Sundrum
``Planck brane''~\cite{LisaRaman2} and/or a ``TeV brane''~\cite{LisaRaman1},
which chop off parts of the
AdS, is very interesting. These constructions have very simple holographic
interpretations, which as we will see illuminate many aspects of the
physics and phenomenology of these models. It is also nice that
 the addition of the Planck and/or TeV branes act as regulators,
and allows for a more intuitive understanding of the holographic
equivalence itself. For instance, adding a ``Planck brane'' to the
gravity side, and making the four dimensional graviton dynamical
by integrating over the boundary conditions,
 corresponds to
putting a UV cutoff $\stackrel{<}{\sim} M_{Pl}$ on the CFT and adding 4D
gravity. Now, the field theory is still conformal and so there is
still no spectrum or S-matrix, but we now have gravity to act as a
probe on both sides of the correspondence. We can now ask the same
physical questions on both sides, such as what is the the leading
correction to Newton's law between two masses on the Planck brane?
On the gravity side, this calculation is weakly coupled, and comes
from the exchange of the continuum of KK gravitons. On the CFT
side, we compute the CFT loop correction to the 4D graviton
propagator, which is a strongly coupled calculation but is
fortunately totally determined by conformal invariance. The fact
that the force law is identical whether one thinks in terms of
living on a Planck brane in an infinite fifth dimension or being
gravitationally coupled to a CFT in 4D is a strikingly physical
illustration of holography. The situation becomes even clearer
with the addition of a ``TeV brane'', which corresponds to some
deformation of the CFT leading to a breakdown of conformal
invariance in the IR. Now, we really do have particles and
S-matrix elements, so the statement of the holographic equivalence
between the broken CFT and the gravitational picture  is the
familiar one, namely
 the two theories
have identical spectra, identical S-matrices.

\section{The Holographic Intepretation of RS2}

Throughout this paper, we will assume that the  AdS/CFT correspondence
can be extended to tell us that
{\it any} 5D gravity
theory on $AdS_5$ is holographically
dual to {\it some} strongly coupled, possibly large $N$,  4D CFT.
We will see that this assumption satisfies various consistency
checks, at least qualitatively in the course of this paper.
In Poincar\'e co-ordinates,
the metric of $AdS_5$ is
\beq
ds^2 = \frac{L^2}{z^2} (\eta_{\mu \nu} dx^\mu dx^\nu + dz^2).
\eeq{n2}
Notice that a rescaling $z \to \lambda z, x^\mu \to \lambda x^\mu$
leaves the metric invariant; translations in $\log z$ correspond to 4D scale
transformations. Loosely speaking, the fifth dimension $z$
is encoded in the 4D theory as the RG scale, with larger $z/L$ corresponding
to the infrared in the CFT.
The self-similarity of the 5D
background is then interpreted as the conformality of the 4D theory.

Beginning with this correspondence, we can start deforming one side in some
way and ask what the deformation means on the other side.
We can start by modifying the gravity side. One thing we can do is add the
Randall-Sundrum ``Planck brane''.
We will review recent
results indicating that
the duality applies in this case \cite{Ed2,Gubser}.

 A convenient way of describing this
is to have a metric which is AdS for $z>z_c$, and simply reflect
this space about $z_c$ for $z<z_c$. The resulting metric is, after
some obvious variable changes and coordinate rescalings,
\begin{equation}
ds^2 = \frac{1}{(1 + |z/L|)^2} (\eta_{\mu \nu} dx^\mu dx^\nu + dz^2).
\end{equation}
There are clearly discontinuities in the curvature at $z=0$, but
they  are attributed to a delta function brane at $z=0$, with a
specific tension finely tuned to the bulk cosmological constant.
This is the ``Planck brane''. Studying the metric fluctuations
about this background reveals a zero mode graviton localized to
the brane at $z=0$, with 4D Planck mass $M_4$ related to the 5D
Planck mass $M_5$ via
\begin{equation}
M_4^2 \sim M_5^3 L .
\end{equation}
There is also a continuum of KK modes in the spectrum, which however has
small couplings to sources on the Planck brane. The continuum
modes give a correction
to the Newtonian potential between two test masses $m_1,m_2$ separated by a
distance $R$ on the brane of the form
\begin{equation}
V(r) = \frac{m_1 m_2}{M_{4}^2} (\frac{1}{r} + \frac{L^2}{r^3}) .
\end{equation}

What does this deformation correspond to in the gauge theory? We
are modifying the space away from pure AdS at small $z$, which
corresponds to the ultraviolet in the gauge theory. The most
natural guess is that we have taken the CFT, put some sort of UV
cut-off near $M_{4}$, and added 4D gravity to the theory. The
precise structure of the ``Planck brane'' is reflected in the
precise nature of the UV cut-off in the dual theory. So, the
deformation of AdS/CFT to the present case seems to suggest that
{\it the RS2 theory with localized gravity in an infinite fifth
dimension is dual to a strongly coupled CFT with a cutoff + 4D
gravity}~\cite{Ed2,Gubser}. This conjecture passes one immediate
check.  From the 4D viewpoint, the leading correction to the
Newton potential comes from the 1-loop correction to the graviton
propagator, which is schematically, in momentum space
\begin{equation}
\frac{1}{p^2} \langle T(p) T(-p) \rangle \frac{1}{p^2} ,
\end{equation}
where $T$ is the CFT energy momentum tensor. Despite the fact that the theory
is strongly coupled, this $\langle TT \rangle$
correlation function is fully determined by conformal invariance
\begin{equation}
\langle T(x) T(0) \rangle \sim \frac{c}{x^8} \to \langle T(p) T(-p) \rangle
= c p^4 \log p^2 ,
\end{equation}
where $c$ is the central charge. When transformed back into position
space, this gives the correct $1/r^3$ correction found from the gravity
side~\cite{Ed2,Gubser}.

If there are fields (say the SM) localized on the Planck brane, then in the
4D theory they interact with the CFT both through 4D gravity, and potentially
also through higher-dimension operators suppressed by $M_4$.

We could have of course started instead from the 4D side, and
deformed the theory by gauging some of its global symmetries. The
most natural symmetry to gauge is the Poincar\'e symmetry of the
4D theory, i.e. add 4D gravity with some Planck scale $M_4$. Since
this corresponds to a big modification at high energies $\sim
M_4$, on the dual side we should expect some breakdown of AdS at
small $z$, and this is the ``Planck brane''. Actually, we need to
be a little more precise about ``cutting off'' the CFT. 
It is important to recognize that from the weakly coupled gravity picture,
we see that the field theory remains conformal in the IR in the presence of
the cut-off and gravity. Obviously a random way of cutting off the CFT and 
adding gravity will not guarantee that the low energy theory is conformal, for 
instance mass terms may be generated that destroy the conformal invariance. 
Consider as an example weakly coupled ${\cal N}=4$ SYM, and add
ordinary 4D gravity (not supergravity). Then, at two loop order,
the 6 scalars pick up quadratically divergent masses near the
cutoff $M_4$, and the low energy theory is certainly not
conformal. Had we added even just ${\cal N}=1$ supergravity, the scalar
masses would be protected and the low energy theory would continue
to be conformal. What is special about the way that conformal invariance 
is broken in the holographic dual of RS2 is that the breaking of conformal
invariance shows up only in Planck {\em suppressed}
operators, not in mass terms. If we want a completely
non-supersymmetric example, we need a conformal theory where, even
in the presence of gravity, there are no relevant deformations
which can not be prohibited by symmetries. For instance, a purely
fermionic conformal gauge theory  with fermion masses prohibited
by chiral symmetries might have this property.

We must remark that the holographic correspondence in the way described
above is only valid at energies $E \ll L^{-1}$. For instance, on the
gravity side,
at distances $r \ll L$, the  gravitational potential turns over to the 5D
$(1/M_5^3 r^2)$ form; the CFT +4D gravity can not possibly reproduce this
result. This is expected from the full duality between the CFT and
string theory on $AdS_5 \times S_5$,
the 5D gravity/4D CFT correspondence is only valid at distances larger than
the radius of the $S_5$ which is $L$.

\subsection{Wilsonian Renormalization and Holography}
Moving the position of the UV brane from $\Lambda=1/L$ to
$\Lambda'<\Lambda$ corresponds to integrating out the degrees of
freedom with energy in between the two cutoffs. This
correspondence has been analyzed in several
papers~\cite{PS,BK,dBVV,VV}. Specifically, ref.~\cite{VV} analyzes
the case where $M_{Pl}<\infty$, where gravity is dynamical. In this
subsection, therefore, we will be brief and only point out to one
simple check of this correspondence. In the RS background, using
the metric in Eq.~(\ref{n2}), the zero-mode of the 5-d graviton
reads \beq g_{55}(z,x)={L^2\over z^2}, \;\;\;
g_{5\mu}(z,x)=0,\;\;\; g_{\mu\nu}(z,x)= {L^2\over z^2}
\gamma_{\mu\nu}(x). \eeq{w1} We set the UV brane at $z_c=L$. The
4-d Planck mass is obtained by substituting the zero mode into the
5-d Einstein action \bea S&=&M_5^3\int_{L}^\infty dz\int d^4x
\sqrt{-g} R^{(5)}(g)= M_5^3\int_{L}^\infty dz {L^3\over z^3} \int
d^4x
\sqrt{-\gamma}R^{(4)}(\gamma)\nonumber \\
&=& { M_5^3L \over 2} \int d^4x \sqrt{-\gamma}R^{(4)}(\gamma).
\eea{w2} This equation gives the result mentioned above:
$M^2_4=M^3_5L/2$. If one adds an IR brane at $z=1/\mu \gg L$, as
in RS1~\cite{LisaRaman1}, then the 4-d Planck constant is: \beq
M_4^2=M_5^3\int_L^{1/\mu}dz{L^3\over z^3} = M_5^3L^3\left( {1\over
L^2}- \mu^2\right) . \eeq{w3} These results for the 4-d Planck
mass are consistent with the interpretation of RS2 as a CFT with
cutoff $1/L$ coupled to gravity, and also suggest that the
holographic dual of RS1 is a CFT broken at scale $\mu$ coupled to
gravity. Both in RS1 and RS2 the Planck scale is fully induced by the
(broken) CFT. Notice that both the scale dependence in
Eq.~(\ref{w3}) and the numerical coefficient $M_5^3L^3$ are
consistent with this interpretation. Indeed, in field theory, the
inverse Newton constant induced by a CFT broken at scale $\mu$ and
with cutoff $1/L$ is $c(L^{-2} -\mu^2)$, where $c$ is uniquely
determined by the central charge of the CFT; the holographic
computation of $c$ using 5-d supergravity gives precisely
$M_5^3L^3$~\cite{HS}.

Generically, quantum gravity may not be fully induced by the CFT.
This happens when a 4-d Einstein action is introduced as a
boundary term on the UV brane, i.e. when the term $M_0^2\int d^4x
\sqrt{-g} R^{(4)}(g)|_{z=L}$ is added to the 5-d Einstein action.  In
this case, the 4-d Planck scale is given by \beq M_4^2=M_0^2 +
M_5^3L^3(\Lambda^2-\mu^2), \;\;\; \Lambda={1\over L}. \eeq{w4}
This boundary term is obviously necessary if we want to integrate
out UV degrees of freedom by moving the Planck brane from $z=L$ to
$z=1/\Lambda'>L$. To keep the physical 4-d Planck mass invariant
we must change $M_0^2$ to $M_0^2 +M_5^3L^3(\Lambda^2-\Lambda'^2)$.
In \cite{Verlinde} for example, where there is an embedding of the
RS-like background into string theory, $M_0^2$ does not vanish.
Sending $M_0 \to \infty$ leaves us with the 5D holographic dual 
of just a cut-off CFT, {\it without} gravity.    

\subsection{Qualitative Holography}
Holography tells us that the $z$ direction of gravity theory can
be interpreted as the RG scale of the 4D theory. It is then
amusing to consider what various ``bulk'' phenomena look like in
the 4D picture. Of course this topic has been extensively
investigated in the infinite AdS case and has also been nicely
analyzed in the RS case \cite{giddingskatz}, so we will briefly
consider one bulk process by way of illustration. Imagine a
graviton, starting at a given point $(x,z_0)$ and aimed straight
at the Planck brane where the SM fields live. After the time
required to traverse the distance to the Planck brane (which in
this background metric is $T \sim z_0$), the graviton bangs into
it and creates e.g. photons. For times $t<T$ the graviton has not
hit the Planck brane yet and so there is no photon production in
the classical limit. How do we understand this from the 4D side?
By holography, a point particle localized at some point $(x,z)$ in
the bulk must correspond to a shell-like configuration of the CFT
of size $z$ centered around $x$. If the bulk particle is moving at
the speed of light towards the Planck brane, from the 4D side the
size of the CFT lump is decreasing at the speed of light. When the
shell is of size $M_4^{-1}$, the standard model fields will couple
to the conformal field theory modes via the gravitational coupling.
Photon production then happens after the time it takes for the
lump to decrease from size $z_0$ to $M_4^{-1}$, which is just
$\sim z_0$, in agreement with the 5D picture.

\subsection{The Strong Coupling ``Problem''}
Purely from the gravity side, the continuum KK modes of RS2 are somewhat
mysterious. While we see them upon
linearizing
 about the background at quadratic order, their self-interactions are
divergent due to the strong coupling region $z \to \infty$. At
large $z$, it is not clear that we can think of the KK modes as
the physical particles in the theory because they are not weakly
coupled. But on the other hand, quantities which inclusively sum
over the KK modes, such as the correction to the Newton potential,
seem perfectly sensible. {}From the holographic point of view,
this is obvious: there are no ``particles'' in the CFT. However,
inclusive questions involving an external probe (like 4D gravity
in this case) can give perfectly sensible answers.

That there is no strong coupling problem can be understood
directly from the gravity side from the argument in
\cite{LisaRaman2}. In order to most clearly understand the
technical issues, it is most convenient to work in momentum space
for 4D and position space in $z$. For a process on the Planck
brane to probe strong coupling in the bulk, we need to use the
brane-bulk propagator $P(p^2_4,z)$. The crucial point is that for
large enough $z$,
\begin{equation}
P(p^2_4,z) \sim e^{i \sqrt{p_4^2} z} .
\end{equation}
All amplitudes involve integrals of propagators convoluted with
$z$ dependent couplings in the bulk, but the couplings never grow
worse than powers of $z$, and therefore the rapid exponential
die-off or oscillation of $p$ ensures that the contributions from
regions with $z > 1/p_4$ are negligible. Processes with external
momenta $p_4$ on the brane are insensitive to what happens at $z >
(1/p_4)$ in the bulk. This obliterates the strong coupling concern
in RS2. It tells us more; even if the geometry were to change to a
different AdS or deviate from AdS altogether for $z$ bigger than
some $z_c$, for momenta $p_4 > 1/z_c$ on the Planck brane we would
not detect measurable physical effects.

All of this has a natural holographic interpretation: external
graviton momenta $\sim p_4$ are only sensitive to the behavior of
the theory at energies $\sim p_4$ and are in particular
insensitive to anything that happens at far lower energies; e.g.
masses can be neglected at very high energies.

\section{Holography and RS1}

\subsection{Generalities}
Putting a ``TeV brane'' into the theory represents a departure
from AdS at large $z$ and must therefore correspond to a breakdown
of conformal invariance in the IR in the 4D side. Notice that the
conformal field theory is badly broken in the
IR; all remnant of conformal invariance disappears beneath this
scale. Furthermore, the low-energy field theory is weakly coupled and hence
should not have a weakly coupled gravity description.
This holographic description  is consistent with the form of the
induced Newton constant we obtained earlier in Eq.~(\ref{w3}). Now
that the conformal invariance has been broken, we expect to get
physical particles that we can asymptotically separate and so we
can have S-matrices. So, the statement of the holographic
equivalence is simply the intuitive one; the spectra are identical
and the S-matrices are the same! In particular, the KK gravitons
in the gravity side can be interpreted in the 4D theory as
resonances. Of course, since the 4D theory is strongly coupled we
could have never guessed that the low-lying spectrum of
excitations includes a tower of spin-2 particles. For this, the
gravity description is more useful, at least over the energy range
where it is weakly coupled.  In order for the gravity description
to be weak, we need that $L M_5
> 1$; note that
all the KK graviton masses are quantized in units of $\mu $ and
they become more and more narrow as $M_5 L$ increases. So, if we
live on the TeV brane, the 5D gravity description is weakly
coupled, and gets strongly coupled for $E > \tilde{M}_5\equiv
M_5\mu L$. Actually, since the first KK modes have mass $\sim \mu
$, the theory does not look 5D until $E>\mu $, so that the 5D
gravity picture is useful in the regime $\mu<E<\tilde{M}_5$. In
\cite{LisaRaman1}, $M_5 L$ is chosen to be somewhat bigger than
one to for a reliable weak gravity description  (since the
hierarchy was being generated by the exponential warp factor there
was no need to make $M_5 L$ very large).
 We can however
comfortably imagine $M_5 L$ to be as big as one or two orders of magnitude.

In RS1, the SM fields live on the TeV brane. We now consider the
question of  how they to be interpreted on the 4D side. In this
analysis, we will assume there is a valid description at energies
above the TeV scale. In this case,  since from the 5D side we see
that the SM states become strongly coupled to the KK gravitons
above $\tilde{M}_5$, and since from the 4D side these KK gravitons
are CFT bound states,
 we cannot think of the SM states as ``spectators'' to the strong dynamics.
Rather, the SM must emerge as bound states out of the strong
dynamics. So, RS1 does not literally correspond to a technicolor
theory (we will see what does when we discuss gauge fields in the
bulk). Rather, it is a theory where the SM becomes embedded in a
strong CFT above the TeV scale. As such, it shares features in
common with the proposal of \cite{VF} to solve the hierarchy
problem by embedding the SM in a conformal theory at a TeV. There
are some important differences, however: the models of \cite{VF}
were not at strong coupling, and furthermore the explicit non-SUSY
models they considered were obtained by orbifolding conformal SUSY
models, and were therefore only conformal in the large $N$ limit,
requiring a huge $N \sim 10^{30}$ to stabilize the weak scale
\cite{CST}. As discussed in the introduction, it is a nontrivial
thing to break conformal symmetry while not reintroducing
quadratic divergences. The cutoff of RS1 must do this, as is
readily seen from the gravity side. Generic cutoff procedures will
not work.

Finally, an exercise in qualitative holography very similar to what
we did for RS2 is instructive. Suppose we aim a graviton from e.g.
the Planck brane towards the TeV brane. From the gravity point of view, it
takes some time $\sim$ TeV$^{-1}$ for the
graviton to reach the TeV brane and produce
SM particles.
{}From the CFT point of view, we have a spherically symmetric lump of CFT that
starts with a small size and grows at the speed of light.
Only when it reaches a size
of order TeV$^{-1}$, however, can this lump know about the breaking of the
CFT and create SM particles, and this takes the same time TeV$^{-1}$ as
in the gravity picture.

\subsection{``String Theory'' at a TeV}
Of course, all of this physics is in principle contained in the
strongly coupled field theory, and for many processes involving
energies far above $\tilde{M}_5$, in the 4D picture the breaking
of conformal invariance is unimportant and thinking in terms of a
strongly coupled CFT is most useful. For instance, since we see
that the theory gets strongly coupled at $\tilde{M}_5$ from the 5D
side, we would conclude that we hit ``string theory'' at
$\tilde{M}_5$. This is certainly true; we hit string theory
propagating on this particular AdS background. It's just that
string theory on this background behaves quite differently from
what we normally think of as string theory in flat space; AdS/CFT
tells us that this string theory is really the ``QCD string'' of a
strongly coupled 4D CFT. Let us try to better understand physics
at energies above $\tilde{M}_5$. For instance, what happens as we
heat the system to temperatures $T \gg \tilde{M}_5$? One might
think, from a naive expectation of string theory, that there would
be a Hagedorn limiting temperature. However, just as in QCD, no
such limiting temperature exists. At high energies the theory is
described by a (strongly coupled) 4D CFT, and despite the strong
coupling, conformal invariance is enough to tell us the free
energy is the usual one for radiation $F \sim T^4$. Why do not we
get the Hagedorn limiting temperature? After all, from the 5D
viewpoint, we certainly have massive string modes in the bulk, and
their wave functions will be localized on the TeV brane. We may
also have open strings stuck to the TeV brane. So why do not we get
the usual Hagedorn exponential density of states? In the analogous
case in QCD, what happens is that the hadrons become so broad that
they cannot be thought of as particles any more, and instead we
see quarks and gluons at high temperature. Exactly the same thing
happens here. These string modes are just broad bound states of
the CFT, and do not give rise to a Hagedorn spectrum of physical
states.

It is not surprising that a strongly coupled 4D theory should have
broad resonances; what is more interesting is that we get narrow
resonances: the spin 2 KK gravitons! Note that the KK gravitons
are narrow when $M_5 L$ is large, which corresponds to large $N$
in the 4D theory, so this is perhaps not unexpected. We get
roughly $M_5 L$ of these resonances starting at $\mu$ and ending
at $\tilde{M}_5$, above which they too become too broad to be
called particles. So, following the change in free energy as the
system is raised first above temperature $\mu$ to just beneath
$\tilde{M}_5$, we have
\begin{equation}
F^{-}(T) = T^5 \mu^{-1}, \qquad \mu <T<\tilde{M}_5.
\end{equation}
Note again that in this intermediate region $\mu < T <
\tilde{M}_5$, the free energy looks that of a 5D theory. However,
for temperatures $T >\tilde{M}_5$, the free energy should be same
as the pure CFT at temperature $T$, which is $F^{+}(T)\sim c T^4$
where $c$ is the central charge, $c = (M_5 L)^3\equiv
\tilde{M}_5^3 \mu^{-3}$~\cite{gkp2}. Note that there is a
transition in $F$ between $T<\tilde{M}_5$ and $T>\tilde{M}_5$:
\begin{equation}
F^{-}(\tilde{M}_5) = \tilde{M}_5^4 (M_5 L), F^{+}(\tilde{M}_5) =
\tilde{M}_5^4 (M_5 L)^3.
\end{equation}
The difference $F^{+}(\tilde{M}_5) - F^{-}(\tilde{M}_5)$
signals a deconfining phase
transition in the broken CFT at $T=\tilde{M}_5$.

\subsection{Black Holes at a TeV}
One of the features we expect of
a quantum gravity theory is that, when particles are
scattered at energies high above the Planck scale, semi-classical black-holes
are formed. Once they are made, these black holes decay via Hawking radiation.
In a usual flat space background, for $E \gg M_{Pl}$, as long as
the impact parameter of the collision is less than the Schwarzschild radius
$R_E$ of the would be black hole with the c.o.m. energy $E$,
a black hole will form with essentially unit probability.
The production cross-section is then simply the geometrical area
\begin{equation}
\sigma(E) \sim R(E)^2 .
\end{equation}
In 4 flat dimensions, $R(E) \sim E/M_{Pl}^2$, so $\sigma
\sim E^2/M_{Pl}^4$: the cross-section grows indefinitely.

What is the situation in RS1? For the moment, let us assume $M_5L
\sim 1$ for simplicity. Also, since we are interested in creating
5D black holes centered on the TeV brane, let us ignore the Planck
brane, which is essential for 4D gravity but has negligible effect
for 5D black holes. Of course, whatever 5D black hole production
and decay is, it corresponds to some physics in a strongly-coupled
4D field theory, but once again the real question is a physical
one: what do we see when we scatter SM particles at energies $E
\gg$ TeV? Recall that from the broken CFT point of view, the SM
particles are bound states of size TeV$^{-1}$. Scattering SM
particle at energies $\gg$ TeV is much like scattering protons at
energies $\gg \Lambda_{QCD}$. If the impact parameter is bigger
than TeV$^{-1}$, nothing happens! This is already a dramatic
difference from the usual flat space intuition, but it is equally
evident from the gravity side. Recall that in the flat space case,
the only reason that two particles at high impact parameter can
form a black hole is that they attract each other through the
long-range gravitational force. However, the long-range gravity
between two particles on the TeV brane is exponentially small, due
to the gap in the KK spectrum. So, while it is true that a mass
of, say, 10,000 TeV would form a black hole of Schwarzschild
radius 100 TeV$^{-1}$ centered on the TeV brane, two particles of
energy $5000$ TeV impinging on each other with this large an
impact parameter would simply fly past each other without forming
a black hole.  Only if the impact parameter is less than $\sim$
TeV$^{-1}$ can they collide and form the black hole, but the
cross-section for doing this never gets bigger than $\sim$
TeV$^{-2}$.

On the other hand, if we consider the case $M_5 L\equiv
\tilde{M}_5\mu^{-1} \gg 1$, say $\sim 10$, then black holes of
size between $\tilde{M}_5^{-1}$ and $\mu^{-1}$ are essentially the
flat space black holes, and since they are smaller than
$\mu^{-1}$, the gap is irrelevant and they are made with the usual
flat space cross section, which grows with energy. Notice that in
all cases, the black holes are some excitation of the CFT and,
since the CFT is broken, will eventually decay to the lightest CFT
bound states, which are the SM particles. To the extent that $M_5
L$ is large, this should mimic the thermal radiation one expects
from the 5D gravity picture.

\subsection{The Radion}
In the original RS1 model, by fine-tuning the tension of the
negative tension brane, a static background geometry was obtained,
with the separation between the branes determined by the
expectation value of a modulus. This ``radion'' degree of freedom
can be identified with perturbations of the metric of the form
\begin{equation}
ds^2 = \frac{1}{(1 + |z|/L)^2}[dx_4^2 + T^2(x) dz^2].
\end{equation}
Since all mass scales on the negative tension brane are set by $T$,
the radion couples conformally to all states on the IR brane.

The radion wave function is peaked on the TeV brane, and all of
its couplings are 1/TeV suppressed~\cite{LisaRaman1}; indeed this
mode survives even if the Planck brane is removed to infinity. As
such, it must be an excitation of the dual 4D theory. But how can
the scale of conformality breaking in the dual theory be
undetermined? We are not used to this in non-SUSY theories, but
SUSY theories almost always have moduli spaces, and if they are
conformal they are usually only conformal at the origin of moduli
space. Moving along the moduli space then gives a soft breaking of
conformality.

In any case, given that the radion exists even in the absence of
the Planck brane, we can identify it in standard AdS/CFT examples.
For instance, take ${\cal N}=4$ SYM, and move along the Coulomb
branch in such a way as to break the nonabelian symmetry
completely. Now, on the gravity side, moving along the Coulomb
branch causes a departure from AdS at large $z$, and the 5D metric
develops a naked singularity. This singular region plays the role
of the TeV brane in this case. Examining the perturbations on the
gravity side indeed reveals an exactly massless ``radion'' with
wave function peaked on the TeV brane.

The corresponding CFT mode
is easily identified. Among the moduli in the ${\cal N}=4$ theory,
one modulus $T$ can be taken to set the overall scale of all
the adjoint vevs, and the others can be taken to be dimensionless. Since
this is the only source of conformal violation in the theory, every mass
parameter will be proportional to $T$, which is just the same as the conformal
coupling of the radion.

Now we can discuss the holographic interpretation of the
Goldberger-Wise stabilization mechanism \cite{GW}. 
They consider a bulk
scalar field $\phi$ of mass $m^2$, with $(m^2L^2)$ somewhat small
(throughout the rest of this section, we will work in units where
$L=1$). They also put large potentials on the Planck and TeV
branes that energetically force $\phi=v_{UV}$ on the UV brane and
$\phi = v_{IR}$ on the TeV brane with $v_{UV} \neq v_{IR}$. 
(The presence of a large potential for $\phi$ on the branes
also gives a large positive mass$^2$ to the 
zero mode of $\phi$ in the 4D theory, regardless of the sign of $m^2$). 
Minimizing the energy stored in $\phi$ in the bulk then fixes the
interbrane separation. In detail, the bulk scalar equation of
motion is (neglecting the back-reaction of $\phi$ on the metric)
\begin{equation}
(\partial^2 + m^2) \phi = 0 \to z^2 \phi'' - 3 z \phi' - m^2 \phi = 0.
\end{equation}
A trial solution of the form
\begin{equation}
\phi(z) \sim z^{\Delta},
\end{equation}
gives a solution as long as $\Delta$ satisfies
\begin{equation}
\Delta ( \Delta - 4) = m^2.
\end{equation}
The most general solution for $\phi$ in the bulk is then of the form
\begin{equation}
\phi(z) = A z^{\Delta_-} + B z^{\Delta_+}
\end{equation}
where $A,B$ are constants that will have natural interpretations in the
holographic picture. Fixing $\phi(z=z_{UV}=1) = v_{UV}$ and
$\phi(z=z_{IR}) = v_{IR}$ then allows us to solve for $A,B$.
When $m^2$ is small and $z_{IR}$ is large, we can approximate
$\Delta_+=4 + \frac{m^2}{4}, \Delta_- = -\frac{m^2}{4}$
and
\begin{equation}
A = v_{UV}, \qquad B = \frac{1}{z_{IR}^4} \left(v_{IR} - v_{UV}
z_{IR}^{-\frac{m^2}{4}} \right).
\end{equation}
The energy stored in the $\phi$ field
\begin{equation}
V(z_{IR})= \int_{1}^{z_{IR}}dz \frac{1}{z^5} \left(z^2 \phi'^2 + m^2
\phi^2 \right)
\end{equation}
is easily computed; the leading contribution in the expansion in $m^2$ is
simply
\begin{equation}
V(z_{IR}) = 4{z_{IR}^4} B^2  .
\end{equation}
This potential is minimized when $B = 0$, which determines
$z_{IR}$ via
\begin{equation}
v_{IR} = v_{UV} z_{IR}^{-\frac{m^2}{4}} \to z_{IR} =
\left(\frac{v_{IR}}{v_{UV}}\right)^{-\frac{4}{m^2}}
\end{equation}
and it is evident that an exponential hierarchy can easily be generated.
Note that the GW mechanism works whether $m^2$ is
positive or negative \cite{Freedman}.
In order to generate a large hierarchy, we need
$v_{UV} > v_{IR}$ for $m^2>0$ and $v_{UV} < v_{IR}$ for $m^2 < 0$.

Now for the holographic interpretation. In the 4D picture, the
bulk scalar field $\phi$ is associated with perturbing the CFT
by the addition of an operator ${\cal O}_\phi$, of dimension $\Delta_+$.
The quantity $A z^{\Delta_-}$ is the running
coupling constant for this operator at scale $z$, or in other words at the
UV cutoff, the Lagrangian is perturbed as 
\begin{equation}
{\cal L}_{CFT} \to {\cal L}_{CFT} + A O_{\phi}.
\end{equation}
The quantity $B$ is the vev of the  operator $O_{\phi}$:
\begin{equation}
B = \langle O_{\phi} \rangle.
\end{equation}
When $m^2$ is small, $O_{\phi}$ is marginally relevant (for $m^2<0$) and
marginally irrelevant (for $m^2 > 0$).

Consider first the case where $m^2 <0$. Then, the coupling starts
at $v_{UV}$ and grows slowly in the infrared, until it finally
hits a critical value $v_{IR}$ where it triggers a complete
breakdown of conformality (reflected as the IR brane in the
gravity picture). This is the familiar picture of dimensional
transmutation; a large hierarchy being generated by a marginal
coupling getting strong in the IR. Since $B$ vanishes, in this
vacuum $\langle O_\phi \rangle = 0$. The case $m^2 > 0$ is less
standard, but equivalent in many respects. Here, it is better to
think of generating the Planck brane (in some sense, these two
theories are equivalent by interchanging $z \to 1/z$). In this case,
the coupling starts at  $v_{IR}$ and  increases slowly in the UV,
until when it gets larger than a critical value it triggers the
breakdown of conformality in the UV. Notice for this
interpretation, it is critical that conformality is broken via
mass suppressed operators in the UV, but mass operators in the IR.
Again in this vacuum $\langle O_\phi \rangle = 0$.

We can also see what corresponds to changing the interbrane separation.
Suppose we move the IR brane from $z_{IR}$ to $z$. Then, as long as $z$ 
is still far from the UV brane $z \gg 1$, we see that
\begin{equation}
A = v_{UV}, \qquad B(z) = \frac{v_{IR}}{z^{\Delta_+}} \left(1 -
(\frac{z}{z_{IR}})^{-m^2/4} \right).
\end{equation}
This means that the UV coupling is begin kept fixed, but the theory is in a
state $|\psi_z \rangle$,
different from the vacuum, where $B = \langle \psi_z|O_\phi|\psi_z\rangle
\neq 0$
differs from the vacuum expectation value of $O_{\phi}$, 
$\langle O_{\phi} \rangle=0$.
Holding the IR brane at a position $z$ corresponds to minimizing $\langle
\psi|H_{CFT}|\psi\rangle$ over all states $|\psi\rangle$ subject 
to the constraint that
$\langle\psi|O_{\phi}|\psi\rangle = B(z)$. 
Following standard arguments, this leads us
to identify the radion with the Legendre transform of the source for
$O_{\phi}$, or put another way, the radion is the interpolating field for the
operator $O_{\phi}$. We recall this standard argument here for completeness.
Suppose we want to minimize $\langle \psi| H |\psi \rangle$
subject to the constraint that
$\langle \psi| O| \psi \rangle = \phi$. We can do this by introducing
a Lagrange multiplier $J$
and minimize
\begin{equation}
\mbox{min}_{|\psi \rangle, J} \langle \psi |H| \psi \rangle -
J (\langle \psi| O |\psi \rangle - \phi)
= \mbox{min}_{|\psi\rangle,J} \langle \psi |(H - J O)|\psi \rangle + J \phi.
\end{equation}
Let first minimize w.r.t. $|\psi \rangle$, which only enters the first term.
This is minimized when $|\psi \rangle = |O_J\rangle$, which is the ground
state of the Hamiltonian $H_J = H - J O$. The ground state energy is
nothing other than $W(J)$. So, now, we have to minimize
\begin{equation}
\mbox{min}_{J} \left(W(J) + J \phi\right).
\end{equation}
The minimum occurs at some value for the source $J_\phi$, and the value of the
function at the minimum is nothing
other than the Legendre Transform $\Gamma(\phi)$.
So, we have learned that the minimum value of the energy $\langle H\rangle$ 
subject
to the constraint $\langle O\rangle = \phi$ is 
$\Gamma(\phi)$, and the state for which this
minimum is attained is $|0_{J_\phi}\rangle$, which is the {\it ground} state
of a {\it different} Hamiltonian $(H - J_\phi O)$.

\subsection{4D Quantum Gravity at $M_{Pl}$}
We have understood from both the 5D gravity and 4D broken CFT
points of view that there is interesting strongly coupled physics
at the TeV scale in RS1. However, it is equally apparent from both
sides that 4D quantum gravity does not become important until we
reach energy scales of order $M_{Pl}$. So, all the physical
questions involving strong 4D gravity, such as the resolution of
4D black-hole singularities or the big-bang singularity, can only
be answered at $M_{Pl}$. This runs counter to the naive
expectation from the 5D picture that the ``strings'' we see at the
TeV brane are just red-shifted from the ``strings'' we would see
on the Planck brane, which would lead us to conclude that we {\it
can} learn about 4D quantum gravity at $M_{Pl}$ by probing the
theory at a TeV. However, this intuition is incorrect because
although there do exist string modes localized to the TeV brane,
just as the zero mode graviton is localized on the Planck brane,
with exponentially small wave function on the TeV brane, all the
rest of the states responsible for making 4D gravity finite are
also localized on the Planck brane and are therefore inaccessible
at TeV energies. This point is also quite clear in the string
realization of these models along the lines of \cite{Verlinde}.
There, it is true that the QCD strings and the strings responsible
for making gravity finite are in fact the same type IIB string.
However, physics near the TeV and Planck branes probe the behavior
of this IIB string on very different backgrounds. Close to the TeV
brane, we are probing IIB string theory on the (deformed by the IR
brane) AdS geometry, where it is indistinguishable from a QCD
string. On the Planck brane, on the other hand, we are probing the
IIB string on an essentially flat (compact) background, where it
does {\it not} look like a QCD string, but the more conventional
string theory in flat space.

\subsection{Effective field theory}

Finally we wish to make a comment on the effective field theory of
these warped geometries. From the gravity side, at every $z$ the
theory is weakly coupled beneath the local cutoff $M_5(L/z)$ (in a 
string theoretic setting, the local cutoff would actually be 
$M_s (L/z)$, where $M_s$ is the string scale). 
Furthermore, because of the redshift factor, all processes which
start within the domain of validity of this effective theory
remain within the effective theory. So we are free to write down
any sort of effective theory we wish; for instance we can put SM
fields on the TeV brane, or move the fermions around in different
places in the bulk etc. However, not every effective theory can be
consistently embedded into a full theory that has  a sensible
interpretation above a TeV. Normally, from a bottom-up point of
view we do not think of this as being too constraining, since the
detailed structure of the underlying ``theory of everything'' is
not known. As long as the low energy theory is consistent, we can
imagine that it will somehow be embeddable into the full theory.
Of course, our lack of knowledge about the full theory means that
we can't even in principle answer questions about physics at
energies far above the cutoff. However, in the case where the
effective theory is that of weakly coupled gravity on the $AdS_5$
geometries, AdS/CFT gives us a non-perturbative definition of the
gravity theory, and tells us that the ``theory of everything''
that the effective theory gets embedded into is a strong 4D CFT.
Therefore, while from the 5D point of view it seems perfectly
innocuous to put SM fields on the TeV brane in the effective
theory of RS1, we understand that embedding this picture into the
fundamental theory requires us to find a broken CFT where the SM
fermion, gauge boson and Higgs fields emerge as composites.

\section{Gauge Fields in the Bulk}

\subsection{Phenomenology}
Several groups have considered putting the SM gauge fields in the
bulk of the RS1 geometry. The phenomenology of these models have a
number of peculiar features quite different from the case of just
gravity in the bulk. For instance, when the brane separation is
chosen to generate the hierarchy, the wave function of the KK
gauge bosons on the Planck brane is sizable, about $.2 \times$ SM
gauge couplings,
 while the KK graviton wave functions are strongly
suppressed there. The KK gauge boson wave function on the TeV
brane are a factor of 8 larger than the gauge coupling, with
consequently severe limits on the mass of the first KK gauge boson
in excess of 10 TeV if the SM fermions live on the TeV brane. In
an effort to circumvent this problem, \cite{GP} considered the SM
fermions localized away from the TeV brane at different points in
the bulk, with the Higgs still on the TeV brane to maintain the
solution to the hierarchy problem. The small overlap between the
wave function of the left/right handed and Higgs fields helps
generate a fermion mass hierarchy as in \cite{AS}
\footnote{Actually in order to avoid FCNC problems from
non-universal couplings to the KK gauge bosons, these KK modes
must either be made heavier than $\sim 100$ TeV or the first two
generations cannot be split, and the first-second generation
hierarchies must come from a separate source \cite{AS}.}

Pomarol \cite{Pomarol} considered the case of an SU(5) gauge
theory broken to the SM in the bulk, with all the SM fields matter
fields on the Planck brane. (Of course this now requires SUSY for
the hierarchy problem, but if SUSY breaking is triggered on the IR
brane then the warp factor can still be used to generate an
exponential hierarchy, with the SUSY breaking transmitted to the
matter fields via the SM gauge interactions.) The zero modes of
the $X,Y$ gauge bosons pick up a mass $\sim M_{GUT}$, while the KK
excitations of the $X,Y$ are at $\sim$ TeV. Furthermore, unlike
the SM KK gauge boson excitations, the $X,Y$ zero and KK modes
have exponentially suppressed wave functions on the TeV brane, and
so give no problems with proton decay. However, since the $X,Y$ KK
modes are charged under the SM, they can be pair-produced at
energies $\sim$ TeV. Pomarol also computed the renormalization of
the SM gauge couplings, finding the interesting result that
despite the presence of the fifth dimension and KK gauge bosons near a TeV, 
the gauge couplings
continue to run logarithmically as in 4D, unifying at  $M_{GUT}$
in the usual way. However, this is not quite true. In fact, we
will see there are large threshold corrections. This can again be
seen on both sides of the duality.

 As we will see, all of the
above scenarios have a very simple holographic interpretation that
greatly clarifies the physics. We summarize the main results here
and present technical details in the next two subsections. The key
point is that holographic dual of having gauge fields in the bulk
with gauge group $G$ is a broken 4D CFT, where a subgroup $G$ of
the global symmetry $G_{gl}$ of the CFT has been weakly gauged.
This is much as in the SM, where $SU(2)_L \times U(1)_Y$ weakly
gauges a subgroup of the $SU(N_f)_L \times SU(N_f)_R \times
U(1)_V$ global symmetry of QCD, with leptons added as spectator
fermions to cancel anomalies. In the simplest set-up, the Landau
pole for the 4D gauge coupling is at the cutoff $1/L$ of the CFT,
and it runs logarithmically to weaker values at lower energies.

Just as KK gravitons are spin 2 bound states
of the broken CFT, the KK gauge bosons are spin 1 bound states.
More specifically,
whatever representation $R$ of $G_{gl}$ the CFT matter fields $Q$ fall
into, there will be  $(Q \bar{Q})$ bound states transforming as
the adjoint of $G_{gl}$ and in particular as adjoints under the $G$ subgroup.
The bound states of this sort which have spin 1 are the KK gauge bosons.
If the SM fermions are taken to live on the TeV brane, then they are composites
of the broken CFT, and therefore their couplings to the spin 1 bound states
are unsuppressed, surviving even in the limit where the SM gauge couplings
are sent to zero. This provides a natural explanation for the stringent
limits on this scenario found in \cite{DHR}.

The dual description of the scenario of \cite{GP}, where the SM
fermions are localized away from the TeV brane, but with the Higgs
localized on the TeV brane, is similar in some respects to``
walking extended technicolor'' \cite{Holdom}. In contrast with
RS1, where the SM gauge bosons must be thought as bound states of
the broken CFT, here we have a situation where the SM is weakly
gauged into a strong sector. Having the Higgs on the TeV brane is
tantamount to saying that the strong sector spontaneously breaks
some of its global symmetries, (the SM subgroup of which are
weakly gauged). But this is exactly what we mean by technicolor.
It is like ``walking'' technicolor because the strong dynamics has
a large window where it is nearly conformal. However, in true
walking technicolor, only some operators scale so slowly, whereas
there is no such distinction here.  Finally, a fermion mode
localized in the bulk has the interpretation of a Higgsing of the
high energy CFT, giving at low energies the fermion and another
CFT. Integrating out the modes that become heavy through the
Higgsing generates higher-dimension operators between the fermions
and the low-energy CFT, and when the low energy CFT finally
triggers electroweak symmetry breaking the fermions acquire their
masses. This is essentially an {\it extended} technicolor theory.
The difference is that the standard model fermions here are
conformal field theory bound states. Furthermore, it is difficult
to see how to generate generational mixing in this way.

The GUT scenario of \cite{Pomarol} also has a simple holographic
interpretation. We have a broken CFT with an SU(5) global
symmetry, the SM subgroup of which is gauged at low energies. (We
could imagine that the full SU(5) was gauged and broken at the GUT
scale, with the real $X,Y$ gauge bosons picking up a mass
$M_{GUT}$.) Despite the fact that the CFT is strongly coupled, the
4D gauge couplings continue to run logarithmically far above the
TeV scale. This is analogous to the fact that the QED gauge
coupling runs logarithmically despite the coupling to quarks which
feel the strong confining dynamics of QCD. As we will see, in the
CFT case we can exactly compute and reproduce the logarithmic
running despite the strong coupling, since the relevant $\langle j
j \rangle$ correlator is fixed by conformal invariance. {}From
this perspective, it would seem that  the unification of couplings
is preserved since the CFT states come in complete SU(5)
multiplets. However, the number of GUT representations is large,
as it is determined by $N$ which is large if we are in the
supergravity regime. So we expect unification will not survive
this scenario. That is, there should be additional threshold
corrections from the gravity picture, which are proportional to
$L/g_5^{2}$. Because $g^2_5$ acts as a cutoff suppressing higher
dimensional operators, one must require $L$ large compared to this
scale, so that we expect large threshold effects.

Since the CFT has an SU(5) global symmetry, there will be spin 1
resonances falling into the {\bf 24} of SU(5). These decompose
under the SM gauge group as adjoints (the SM KK modes), and states
with $X,Y$ gauge boson quantum numbers (the $X,Y$ KK modes). The CFT
picture gives a simple explanation for the puzzling fact that the
SM fermions have large amplitude for producing SM KK modes on the
Planck brane, while the amplitude for producing $X,Y$ KK modes and
KK gravitons is exponentially suppressed. From the CFT point of
view, the SM fermions interact with the CFT either indirectly
through 4D gravity and the SM gauge interactions, or directly via
higher-dimension operators suppressed by $M_{Pl}$. Since all the
direct couplings between the SM fermion and the CFT are very
suppressed, how can we get a sizeble amplitude for producing KK
gauge bosons, which are after all bound states of the CFT? The
answer is these spin 1 bound states are produced through mixing
with the SM gauge bosons. This is like the electroproduction of
the $\rho$ meson in QCD: there is no direct coupling between
electrons and quarks, but the $\rho$ can be produced via its
mixing with the photon, with an amplitude suppressed only by
$e_{QED}$. Similarly in our case, the SM KK gauge bosons can be
produces with an amplitude only suppressed by SM gauge couplings.
On the other hand, the spin 1 bound states with $X,Y$ quantum
numbers {\it can not} mix with the SM gauge bosons because they do
not have the same quantum numbers, and so the amplitude for singly
producing them can only come from $1/M_{Pl}$ effects. They can
however obviously they can be pair produced with SM gauge coupling
strength.

So it appears that one can almost obtain a consistent grand
unification scheme. However, one has to contend with large
threshold effects which seem to  destroy this unification.

\subsection{Localization of Bulk Gauge Fields}
The presence of gauge fields in the 5-d bulk gives rise to new
interesting phenomena.  Some of the material in this section has
also been discussed in~\cite{KSS}, and resulted from discussion
with E. Witten.

In this section the metric is
\beq
ds^2={L^2\over z^2}(\eta_{\mu\nu}dx^\mu dx^\nu +dz^2), \;\;\;
L\leq z \leq 1/\mu.
\eeq{m1}
The Planck brane is located at $z=L$, while  $z=1/\mu$
is the position of the TeV brane: $\mu=O(1\, TeV)$.

By giving Neumann boundary conditions to the gauge field $\hat{A}_m$,
one finds a discrete KK tower of states that begins with a massless mode,
constant in $z$: $\hat{A}_5=0$, $\hat{A}_\mu(z,x)=A_\mu(x)$.
Substituting this zero mode into the canonical action of a 5-d gauge field,
with 5-d gauge coupling $g_5$, one finds
\vskip .1in

\bea
{1\over 4g_5^2}\int_L^{1/\mu} dz \int d^4x \sqrt{-g}g^{mp}g^{nq}
F_{mn}(\hat{A})F_{pq}
(\hat{A})&=&-{1\over 4g_5^2}\int_L^{1/\mu} dz \int d^4x {L\over z}
F_{\mu\nu}^2[A(x)]\nonumber \\
&=&-{L\over 4g_5^2}\log \mu L \int d^4x F_{\mu\nu}^2[A(x)].
\eea{m2}
The 4-d coupling constant that one reads off this formula is
\beq
g^2={g_5^2\over L\log \mu L}.
\eeq{m3}
Since $g$ depends logarithmically on the TeV scale $\mu$ it seems natural to
interpret Eq.~(\ref{m3}) as a running coupling constant evaluated
at the infrared scale $\mu$.
Notice that this is the ``standard'' KK recipe if we identify $\mu$ with
the compactification scale.
Interpreting Eq.~(\ref{m3}) as the IR 4-d coupling constant
is also consistent with --indeed, required by-- holographic duality.

In our case the correct holographic interpretation is that an RS1 background
with a propagating 5-d gauge field is dual to a
4-d CFT broken at the scale $\mu$, coupled to a 4-d gauge field. More
precisely, the gauge field couples to some conserved current of the broken CFT
with running coupling constant $g(p)$. The coupling at $p=\mu$ is given
by Eq.~(\ref{m3}), and it has a Landau pole at $p=O(1/L)$. The beta
function coefficient $b_{CFT}$ is identified as
\begin{equation}
\frac{b_{CFT}}{8 \pi^2} = \frac{L}{g_5^2}
\end{equation}
Of course, below $p=\mu$, the coupling no longer runs.

To check this interpretation we compare the Green function for two sources
located at the boundary $z=L$, computed in~\cite{Pomarol} with the
propagator of the 4-d gauge field $\tilde{A}$, computed using the
holographic duality.

Before doing this, let us remark
that the  5-d gauge field action is non renormalizable, so that it
makes sense only when dimensionful coupling constant $g_5^2$ is smaller than
the cutoff length $L$; therefore, the 5-d description makes sense only when
$L/g_5^2\gg 1$.

\subsection{Computation of the Propagator}

In the case where there the IR brane is removed, it is trivial to compute the
full gauge boson propagator from the CFT side, essentially copying what
was done for the case of just gravity in the bulk. The correction to the
4D gauge boson propagator is schematically
\begin{equation}
\frac{1}{p^2} \langle j(p) j(-p) \rangle \frac{1}{p^2}
\end{equation}
where $j$ is the CFT current coupling to the gauge boson. Despite the strong
coupling, the $p^2$ dependence of the $\langle j j \rangle$ correlator
is fixed by conformal
invariance, since $j$ is a conserved current and therefore has vanishing anomalous dimension. We therefore have
\begin{equation}
\langle j(p) j(-p) \rangle \sim  p^2 \log p^2
\end{equation}
which yields a correction to the gauge boson propagator
\begin{equation}
\frac{\log p^2}{p^2}
\end{equation}
giving precisely the logarithmic  running of the gauge coupling.
This confirms the result from the previous subsection: despite the
appearance that there is no zero mode for the gauge field when the
IR brane is removed (since the zero mode wave function is flat and
the mode is nonnormalizable) , in fact we have a 4D gauge field
coupled to a CFT, causing the gauge coupling to run
logarithmically to zero in the IR.

We can do a precise calculation of the propagator, even in the presence of the
IR brane, using the prescription of \cite{GKP,Ed}.
First of all, we must solve the 5-d equations of motion of the field
$\hat{A}_m$. This is most easily done by choosing the
gauge $\hat{A}_5=0$, and by Fourier-transforming the fields $\hat{A}_\mu$
with respect to the coordinates $x^\mu$.
Defining $\tilde{A}_\mu(z,q)=\int d^4x \exp(iqx)\hat{A}_\mu(z,x)$ we find:
\bea
\tilde{A}_\mu(z,q)&\equiv & C_\mu(q)\tilde{A}(q,z)=
C_\mu(q)[bqzJ_1(qz) + qz Y_1(qz)], \nonumber \\
\partial_z\tilde{A}_\mu(z,q)&\equiv & C_\mu(q)\partial_z \tilde{A}(q,z)=
qC_\mu(q)[bqzJ_0(qz) + qzY_0(qz)].
\eea{m4}
$C_\mu(q)$ and $b$ are constant.
The boundary condition at $z=1/\mu$ is $\partial_z \tilde{A}_\mu(z,q)=0$;
it sets $b=-Y_0(q/\mu)/J_0(q/\mu)$.

At this point, we use the standard holographic correspondence~\cite{GKP,Ed},
summarized in
Eq.~(\ref{n1}), to compute the self energy of the gauge field.
Concretely, we interpret the 5-d gauge field at $z=L$ as a source for
a conserved current of the broken CFT, and its on-shell 5-d action as
the generating functional of the connected Green functions. For the
two-point function we have, with obvious notation
\bea
\langle J_\mu(0) \tilde{J}_\nu (q)\rangle &=&
\left(\eta_{\mu\nu}-{q_\mu q_\nu\over q^2}\right)\Sigma(q); \nonumber\\
\Sigma(q)&=& {1\over g_5^2}\left.{\partial_z \tilde{A}\over \tilde{A}}
\right|_{z=L} \nonumber \\
&=& q{1\over g_5^2}{Y_0(qL) +b J_0(qL)\over Y_1(qL) +b J_1(qL)}
\nonumber \\
&\approx &  q^2 {L\over  g^2_5}\left[\log(qL/2)
+\gamma  -{\pi Y_0(q/\mu)\over 2 J_0(q/\mu)}
\right].
\eea{m5}
The last approximate equality is valid when $qL\ll 1$.

To find the propagator of the dynamical gauge field, we resum the self-energy
in the usual manner. Call $e$ the coupling of $A_\mu$ to the broken CFT.
Define $\langle A_\mu(0)\tilde{A}_\nu(q)\rangle=
(\eta_{\mu\nu}-q_\mu q_\nu/q^2)\Pi(q)$; then
\beq
\Pi^{-1}(q)=[q^2-e^2\Sigma(q)]=q^2 \left\{1-
{e^2 L\over  g^2_5}\left[\log(qL/2)
+\gamma  -{\pi Y_0(q/\mu)\over 2 J_0(q/\mu)}\right]\right\}.
\eeq{m6}

Let us consider at first large Euclidean momenta $q\rightarrow iq$,
$|q|\gg \mu$. In this case, $\pi Y_0(iq/\mu)/ 2 J_0(iq/\mu)\approx i\pi/2$
and the resummed propagator produces a {\em logarithmically running}
coupling constant
\beq
{1\over e_{eff}^2(q)}= {1\over e^2} -{L\over g_5^2}[\log(qL/2) +\gamma].
\eeq{m7}
This result agrees with ref.~\cite{Pomarol} when $e\rightarrow \infty$, i.e.
when the Landau pole is at $2\exp(-\gamma)/L$, i.e. at an energy of
the order of the UV cutoff $1/L$. Notice that the running is logarithmic
at energies well above the KK scale $\mu$. This effect is easy to understand in
the 4-d field theory dual of the RS compactification, since this dual
is a broken CFT weakly coupled to a gauge field.

The explicit computation of the propagator has shown that our
interpretation of the coupling constant given in Eq.~(\ref{m3}) as
an IR coupling, evaluated at the scale $\mu$ is correct. In
particular, even in the cutoff CFT limit $\mu\rightarrow 0$, the
gauge field does not decouple, since only the IR coupling
vanishes, not the propagator at nonzero $q$. True decoupling is
obtained by removing the UV brane, not by removing the IR brane.

Finally, notice that the position of the Landau pole can be changed by adding
a 4-d boundary term $(1/4e_0^2)\int d^4x F_{\mu\nu}^2(A)$ to the 5-d action
in Eq.~(\ref{m2}).
\subsection{KK Gauge Boson Production}
One of the intriguing results of ref.~\cite{Pomarol} is that if on the UV
brane there exist fields charged under the 5-d gauge field, the amplitude for
direct production of KK gauge bosons is non-negligible. Superficially,
this seems in contradiction with the holographic interpretation.
In the dual 4-d field theory, indeed, the KK gauge bosons are bound states of
the broken CFT, and there is no direct coupling between the ``standard model''
fields, living on the UV brane, and the CFT. All interactions between these
two sectors are mediated either by a 4-d graviton or by a 4-d gauge field.
Graviton-mediated interactions are negligible at energies below $1/L$.
On the other hand, kinetic-term mixing between the gauge field and the CFT
spin-1 bound states may (and indeed does) account for Pomarol's result.

First of all, let us find the spectrum of KK excitations. A KK state
obeys Neumann boundary conditions both at $z=1/\mu$ and $z=L$.
This is consistent with the holographic interpretation of KK excitations as
bound states of the broken CFT~\cite{Ed3}.
The b.c. at $z=1/\mu$ gave $b=-Y_0(qL)/J_0(qL)$; the b.c. at $z=L$ implies
\beq
J_0(q/\mu)Y_0(qL) = J_0(qL)Y_0(q/\mu).
\eeq{m8}
For $\mu L\ll 1$, this equation is approximately solved by
$J_0(q/\mu)=0$; therefore,
the spectrum of KK excitations is discrete and quantized in units of $\mu$.

By coupling the broken CFT to the gauge field $A_\mu$, the mass of the
KK excitations/bound states is given by the position of the poles in
the propagator
$\Pi_{\mu\nu}(q)$, and it is shifted from the $e=0$ value by terms $O(e^2)$.
The mass of the $n$-th bound state is then $m_n=\mu j_{(0,n)} + O(e^2)$;
$j_{(0,n)}$ is the $n$-th zero of $J_0$.

The $n$-th spin-1 bound state couples to Planck brane (``Standard
Model'') fields with strength $F$ given by \beq F^2=2m_n
\mbox{Res}\, e^2\Pi(q)|_{q\approx m_n}  . \eeq{m9} The residue is
most easily evaluated by rewriting $e^2\Pi$, using
definition~(\ref{m7}), as \beq e^2 \Pi(|q|)= {e^2_{eff}(|q|)\over
q^2}\left[1 + e_{eff}^2(|q|){\pi L \over 2 g_5^2} {Y_0(q/\mu)\over
J_0(q/\mu)}\right]^{-1}. \eeq{m10} By expanding the expression in
brackets in powers of $e^2_{eff}$, and thanks to the Bessel
function identities $J'_0(j_{(0,n)})=- J_1(j_{(0,n)})$ and
$J_1(j_{(0,n)})Y_0(j_{(0,n)})=2/\pi j_{(0,n)}$ we find that to
order $e_{eff}^2$, the coupling $F$ is \beq F=\sqrt{{ 2L \over
g_5^2}}{e_{eff}^2(m_n)\over (m_n/\mu) J_1(m_n/\mu)}. \eeq{m11}
This formula reduces exactly to the one found by Pomarol
in~\cite{Pomarol} when the Landau pole is set at
$2\exp(-\gamma)/L$, i.e. when $e\rightarrow \infty$.

\section{Probe Branes}
In \cite{intersecting, LR}, models for solving the hierarchy problem
with infinitely
large dimensions were considered, where the SM fields are localized on a
probe 3-brane away from the brane where gravity is localized. Specifically,
in ~\cite{LR},
there is an infinite $AdS_5$ with a 3-brane located
away from the Planck brane where the SM is taken to live, with the warp factor
generating the TeV scale exponentially as in RS1.
Since we have an $AdS$ both to the left and the right of the probe brane,
the 4D picture must be that at low energies we have $SM \times CFT_L$, which
merge into a single $CFT_H$ theory at energies above $\Lambda_{IR} \sim$ TeV.
An example of this scenario is realized if, for instance,
in the Standard ${\cal N}=4$ case we start with $N+1$ branes and remove a single
``probe'' brane away from the remaining stack of $N$ branes. From the gravity
side we will have something that looks identical to LR (except without the
Planck brane). From the 4D viewpoint, we have just Higgsed
$U(N+1) \to U(N) \times U(1)$ at $\Lambda_{IR}$. Now, LR computed the
correction to Newton's law on the probe brane, finding
\begin{equation}
V(r) \sim \frac{m_1 m_2}{M_{4}^2} (\frac{1}{r} + \frac{L^2}{r^3}) + \frac{m_1
m_2}{
TeV^8 r^7},
\end{equation}
where the last term is new and dominates at short enough distances. This can
be understood from the 4D point of view as follows. After integrating out
massive states at the scale $\Lambda_{IR}$, we are left with irrelevant
operators linking the SM fields to the fields of the low-energy $CFT_L$.
An operator allowed by all possible symmetries is
\begin{equation}
\frac{1}{\Lambda_{IR}^4} T_{SM}^{\mu\nu} T_{\mu\nu\, CFT_L} .
\end{equation}
Exchanging the $CFT_L$ states then generates the operator
\begin{equation}
\frac{1}{\Lambda_{IR}^8} T^{\mu\nu}_{SM}(p) \langle T_{\mu\nu\, CFT_L}(-p)
T^{\rho\sigma}_{CFT_L}(p) \rangle
T_{\rho\sigma\, SM}(-p) =
\frac{c p^4 \log p^2}{\Lambda^8_{IR}} T^{\mu\nu}_{SM}(p) T_{\mu\nu\, SM}(-p),
\end{equation}
which precisely yields the $1/(TeV^8 r^7)$ non-relativistic
potential found by LR. Here we have just posited the existence of
the $T^{\mu\nu}_{SM} T_{\mu\nu\, CFT}$ operator, and showed that
it would reproduce the LR result, but in the ${\cal N}=4$ SYM case
we can prove its existence. If we go out along the Coulomb branch,
Higgsing $U(N+1) \to U(N) \times U(1)$, we can integrate out the
heavy states and obtain the low-energy action. Of course, the
theory is strongly coupled, but some quantities, in particular the
$\Tr F^4$ piece of the effective action, are known to be 1-loop
exhausted by a nonrenormalization theorem. These operators are
equivalent to $T^{\mu\nu}_{U(1)} T_{\mu\nu\, SU(N)}$ operators,
and so in this specific case we can explicitly see that they are
generated.

We can also consider the case of bulk gauge fields. Of course, in
this case these  be the SM gauge fields, since we can not
phenomenologically tolerate the gapless CFT states charged under
the SM gauge group. However, we can imagine some other gauge field
in the bulk, say coupled to $B-L$, under which the SM fermions are
charged. 

In the case of pure gravity in the bulk, the cross-section for
KK graviton production was $\sigma \sim E^6/\Lambda_{IR}^8$, which is
the same as the case of $n=6$ large extra dimensions. Indeed, we even expect
all differential cross-sections to be the same as in the $n=6$ case, since
in both cases the cross-section for graviton production is related to the
imaginary part of the graviton exchange diagram, and this graviton exchange
propagator is the same $1/x^8$. When there are gauge fields in the bulk,
however, the cross-section for producing KK gauge bosons is much
larger. The reason is that
we can produce CFT modes directly through $s$ channel production.
The cross-section is
\begin{equation}
\sigma(s) \sim {L\over g_5^2}\frac{e_{eff}^4(s)}{s},
\end{equation}
and since there is no gap, this is deadly phenomenologically unless the 4D
gauge coupling $e_{eff}$ is tiny.
This can be done either if the $\beta$ function
is huge or if a large boundary gauge coupling is added on the Planck brane.

\section{Inflation and Cosmology}

The 4D holographic viewpoint is particularly useful for thinking
about the early universe cosmology of models with $AdS_5$ slices.
The late cosmology of RS2 has been discussed in e.g. \cite{Cos}.
The early cosmology of RS2 with a heated bulk has been addressed
in \cite{Gubser}. Let us turn to RS1. As we have discussed,
contrary to the naive expectation that we cannot extrapolate
cosmology to temperatures above $\sim$ TeV because of ``hitting
string theory at a TeV'', at temperatures higher than TeV the
universe has normal radiation-dominated FRW expansion from a hot
CFT. In the RS2 case, there is a solution of the 5D gravity
equations which has an induced 4D metric on the Planck brane given
by a radiation dominated FRW cosmology, with a retreating horizon
in the bulk \cite{Gubser}. Physically, this corresponds to the
bulk initially heated up near the Planck brane, with the hot
region gradually expanding away and cooling down. The cosmology
also follows from the analysis of \cite{Kraus}, who considered
other potential cosmological solutions in and AdS-Schwarschild
background, and found the brane moves through the static
background in accordance with the expected four-dimensional
behavior.

Now we consider a theory which at zero temperature has an infrared
breaking of the CFT, to generate a TeV brane. We now ask what this
looks like at early times. From the holographic viewpoint, it is
clear that one does not expect to see a TeV brane at temperatures
well above a TeV, since the ultimate breaking of the conformal
field theory applies at temperatures below that scale. So at high
temperatures, there are two possible scenarios. The first is that
the TeV brane simply does not exist at early times; the horizon
shields the region where the TeV brane might exist. As the
temperature drops and the horizon receeds, it eventually uncloaks
the TeV brane, and the true SM degrees of freedom emerge. An
alternative possibility with similar physical consequences is that
a TeV brane exists at the scale associated with the temperature of
the theory. Only when the temperature drops to appropriately low
scale will the brane settle at its true minimum, analogously to
the behavior of other moduli in the early universe. Notice that in
either case, if we distinguish the AdS curvature and the
fundamental Planck scales, we expect a regime between these scales
multiplied by the TeV warp factor in which the theory appears
five, not four dimensional, as the TeV brane settles to its true
minimum.

There has been some concern about whether inflation can be made to
work in RS, while getting large enough density perturbations. The
worry has been that one cannot put the inflaton on the TeV brane,
since its energy density will be too low to generate sufficient
density perturbations, and that putting it on the Planck brane
would make it impossible to directly reheat SM fields. The CFT
viewpoint shows to the contrary that doing inflation is standard;
we just add an inflaton with $1/M_{Pl}$ couplings to the CFT.
After inflating near the GUT scale to generate appropriate $\delta
\rho/\rho$, the inflaton reheats the CFT, which eventually turns
into SM particles when the temperature drops beneath a TeV. The
bulk picture of all this is the following. After inflating, the
inflaton on the Planck brane reheats the bulk in the immediate
vicinity of the Planck brane. This is clear from the KK picture
since it has unsuppressed couplings to the heaviest KK modes. This
reheating generates a horizon in the bulk and starts us off in the
FRW solution of \cite{Gubser}. The ensuing evolution is the same
as described in the previous paragraph, and the SM is eventually
reheated to $T \sim$ TeV. It would be interesting to investigate
this 5D picture in more detail.

Much of the worry about the early universe cosmology in RS1 has
evolved around the light radion. In particular, if we imagine that in the
early universe the IR and UV branes were far displaced from their final
equilibrium values, then the early cosmology will certainly not be standard
FRW expansion. We can understand this from the 4D viewpoint given
the holographic interpretation of the radion we discussed
in the previous subsection. Moving the IR and UV branes say, much closer
to each other, corresponds to putting the broken CFT in a very special,
{\it non-thermal} state. It is not surprising that such a state
should yield unusual cosmology. If the initial condition for
QCD was a coherent state where e.g. $\Tr(F^2) = M_{GUT}^4$
then we would also expect unusual cosmology! However,
we do not usually imagine such an initial state for the universe. Rather,
we imagine that the early universe was in a hot thermal state,
and indeed this can come about through standard inflation and reheating.
So, while it is perfectly fine to consider cosmology with widely varying
radions as a mathematical exercise, this does not correspond to realistic
initial conditions for early cosmology. The more reasonable initial condition
is a thermal one, where from the 5D point of view the TeV brane is dissolved,
and instead there is a horizon away from the Planck brane, giving the
FRW solution of \cite{Gubser}.

The cosmology of the probe brane scenarios is also interesting. From the
CFT point of view, it is clear that the reheating temperature can not
exceed $\sim$ TeV. If this happens, then as the high energy CFT cools down
beneath a TeV, the energy density get divided into the SM and the
low-energy CFT, which is still a large $N$ gauge theory redshifting as
radiation. This would give too many degrees of freedom during
nucleosynthesis. Therefore, we have to imagine that only the SM degrees of
freedom are reheated.  This might be difficult, since the coupling to the
standard model states at high energy corresponded to a coupling to the
full high energy CFT. It is difficult to see how something could decay
into standard model states but not the low-energy CFT states. Here,
however, we will take the most conservative viewpoint and assume that
initially only the SM fields are heated up. We then demand that they do
not heat up the low energy CFT.

With just gravity in the bulk, the low energy CFT is coupled to
the SM most strongly only via the dim. 8 operator $TT$ operator,
and so the cross section for producing the CFT states with the SM
fields at temperature $T$ is $\sigma \sim T^6/\Lambda_{IR}^8$. In
order not to thermalize the CFT, the mean free path $l \sim (n
\sigma v)^{-1}$ should be much larger than the Hubble size $H^{-1}
\sim M_{Pl}/T^2$, and therefore we have the constraint
\begin{equation}
\frac{T^9}{\Lambda_{IR}^8} \lsim \frac{T^2}{M_{Pl}} \to
T \lsim 10 \mbox{GeV}
\left(\frac{\Lambda_{IR}}{\mbox{TeV}}\right)^{8/7} .
\end{equation}
It is interesting that this constraint on the reheating temperature is
similar to the case of large extra dimensions with $n=6$ \cite{ADDlong}.
However the physics is quite different and the constraint on the probe brane
theories are actually somewhat weaker. In the large dimension case, the
KK gravitons are produced through a process of evaporation: since each
individual KK mode only has $1/M_{Pl}$ interactions, while copious amounts
of energy can be lost into many KK modes, once produced each individual one
hardly interacts with the others or with the SM fields except over huge
timescales comparable or longer than the present age of the universe.
Furthermore, they red-shift away as massive particles, so once
they are produced there is a greater danger that they may come to
dominate the energy density of the universe. The most severe constraint
came from the long-lived decays of the massive KK gravitons
to e.g. photons  or $e^+ e^-$
pairs \cite{ADDlong, HallSmith}.
None of these things happen in the probe brane case, since it is clear from
the 4D viewpoint that the energy thrown into the CFT will redshift as
radiation, and so as long as the nucleosynthesis constraint is obeyed
there is no worry about overclosure or late decays.

If there are gauge fields in the bulk, then unless the 4D gauge coupling is
tuned to be tiny (by adding a large boundary gauge coupling on the Planck
brane), the CFT is immediately thermalized with the SM by the gauge
interactions. The CFT and SM thermalize at a temperature when
\begin{equation}
{L\over g_5^2}e_{eff}^4(T) T \sim \frac{T^2}{M_{Pl}} \to
T = {L\over g_5^2}e_{eff}^4(T) M_{Pl}.
\end{equation}
Requiring that this not happen till after BBN where $T \sim$ MeV
places the constraint
\begin{equation}
\left({L\over g_5^2}\right)^{1/4}e_{eff}(\mbox{MeV}) \lsim 10^{-5}.
\end{equation}

\section{RS and ADD}
We summarize with a comparison of the similarities and differences
between the RS1 and ADD large extra dimensions \cite{ADD} scenarios.  
In both theories, one can view
TeV as the fundamental scale for observers on the brane, with the
Planck scale being an induced scale.  In the ADD scenario, that is
the only point of view; the fundamental scale of quantum gravity
is a TeV; the Planck scale is a result of the small coupling of a
graviton spread over a large volume to the fields on the brane. In
the RS1 scenario on the other hand, it is only observers on the
TeV brane who see the TeV scale as the quantum gravity scale. The
shape of the KK modes (and the string modes) is such that the
scale for 5D quantum gravity (which we have seen is equivalent to
a conformal field theory) depends on the location in the fifth
dimension. Nonetheless, if we live on the TeV brane, we see strong
interactions at a TeV, and KK modes at a TeV. However, we do not
see the string modes associated with the 4D Planck scale.

In this sense, RS1 is a theory which has both the TeV scale and
the Planck scale with a desert in between. Of course, the theory
is strongly interacting at the TeV scale. Nonetheless,
cosmologically and otherwise, one can ask questions pertaining to
physics above the TeV scale. Indeed, as we have discussed these
models have a holographic 4D description by a conventional {\em
local} field theory all the way up to the scale $1/L
\stackrel{<}{\sim} M_{Pl}$. This field theory is strongly coupled
and conformal above a TeV. The solution to the hierarchy problem
using the exponential warp factor from the 5D picture is
translated into  dimensional transmutation in the 4D description.

This being said, the strong coupling field theory gives rises to
many interesting physical phenomena near a TeV, and there is a window of
energies, where the 5D gravity
description is weakly coupled, where these are most usefully thought of as
``5D TeV quantum gravity effects''. The window of energies for which this
5D description is useful is given by $1/z < E < M_5L/z$ where
$z$ is the position of the IR brane. Since the mass of the first KK graviton
is given by $M_{KK} = 1/z$, this window can be expressed as
\begin{equation}
M_{KK} < E < (M_5 L) M_{KK}.
\end{equation}
For $E < M_{KK}$, there is a weakly coupled 4D effective theory, while for
$E > (M_5 L) M_{KK}$ the gravity description is strongly coupled and the
only non-perturbative definition of the theory is in terms of the 4D CFT
description. For $M_5 L \sim 1$ this window does not exist, but we can
certainly imagine $M_5 L \sim 10$ so we could have an order of magnitude
in energy where the 5D gravity description is useful. Many interesting
predictions follow from this picture, for instance, there are about
$(M_5 L)$ KK gravitons starting at $M_{KK}$, which are well defined, narrow
resonances.
These are of interpreted as spin 2 bound states of the 4D broken CFT,
but in the absence of putting the 4D theory on a computer their existence
can only be inferred from the weakly coupled 5D gravity picture.
Furthermore, black holes on the TeV brane with size between
and $(M_5 L M_{KK})^{-1}$ and $M_{KK}^{-1}$ are essentially ``flat space''
5D black holes and are made with the usual cross-section. Again, these states
can be interpreted as excitations of the CFT from the 4D viewpoint,
but the properties of these states are most usefully described in the weakly
coupled 5D picture. On the other hand, the cross-section for making black
holes larger than $M_{KK}^{-1}$ stops increasing
and hits a maximum of $\sim$ TeV$^{-2}$. This is very different than the usual
intuition for black hole production in flat space, where the cross-section
grows indefinitely with energy. As we saw, this difference could be
easily understood both from the gravity and the CFT pictures.

We have seen that processes around a TeV can be most conveniently
thought of as arising from 5D quantum gravity effects, but as we
have remarked RS1 is a theory with 4D quantum gravity at the
Planck scale;  therefore many of the effects we usually associate
with ``quantum gravity'' do not manifest themselves at energies
lower than  $M_{Pl}$. For instance, big-bang cosmology is standard
FRW cosmology for temperatures far above a TeV, and what resolves
the big-bang singularity does not manifest itself before $M_{Pl}$.
The resolution of 4D black-hole singularities occurs at $M_{Pl}$.
4D General Relativity is rendered a finite theory due to
additional states that show up at $M_{Pl}$. This is of course
obvious from the 4D description, because we have simply added 4D
GR to a broken strong CFT, but unraveling the dynamics of the CFT
tells us nothing about what eventually makes gravity a sensible
theory above $M_{Pl}$. Therefore, the experimental observation of
the ``5D TeV quantum gravity effects'' would shed no light on the
4D quantum gravity effects discussed above. From the 5D side, this
can be seen because the closed string spectrum of states falls
into two separate sectors, one near a TeV and the other near
$M_{Pl}$.

By contrast,  the large extra dimensions scenario has only a
single quantum gravity scale $\sim$TeV$^{-1}$. {\it All} quantum
gravity effects are associated with the higher-dimensional Planck
scale, that can be as low as $\sim 1-10$ TeV, and all closed
string states are at a TeV. These theories break the standard
rules for building extensions of the SM (and also make the
construction more challenging), since instead of modifying the
theory starting at short distances $\sim$ TeV$^{-1}$, the theory
is modified at distances much {\it larger} than a TeV$^{-1}$ by
the addition of large new dimensions, or equivalently by the
addition of $10^{32}$ new states (the KK gravitons) {\it lighter}
than a TeV. And finally, since the size of the extra dimensions in
the ADD scenario can be as large as a few hundred microns,
corrections to the $1/r$ Newtonian potential start at almost
macroscopic distances, unlike in RS, where they start at almost
Planckian distances $O(L)$.

One can ask what would be the experimental consequence of this
distinction in theories. {\it If} experiments can reach the
quantum gravity scale (which is above the scale
for the KK modes), one would see quite different strong
interactions. In particular, in the ADD scenario, one would see
``conventional" black hole behavior (appropriate for larger
dimensions), whereas in the RS1 scenario, there would only be a
window of energies for which this applies, after which one sees
cross sections most readily described by the 4D conformal field
theory. In the ADD scenario, one would really be probing the
fundamental gravity theory; this is not the case in the warped
scenarios. One interesting consequence is that the LR theory
\cite{LR} would be distinguished from the six large
extra dimension scenario if and only if the quantum gravity
effects are measured.

It should be noted that all the above conclusions about RS1 apply
to theories in which the warp factor generates the hierarchy,
whereas all the conclusions about ADD apply to theories in which
large dimensions, and no warp factor, are responsible for the
hierarchy. Interesting variants of both these have been
considered, and in some sense, interpolate between the two
different behaviors of quantum gravity. For instance, 
\cite{intersecting} considered intersecting brane
scenarios with a large number of extra dimensions. In this case,
one can generate a hierarchy by choosing a low scale of quantum
gravity, or from the warp factor (or a combination of both). If
the low scale of quantum gravity is chosen of order TeV, and our
physical 3-brane is at the intersection of the branes, the
phenomenology reproduces that of large extra dimension scenarios.
On the other hand, if the hierarchy is produced by placing the
3-brane somewhere  away from the intersection, then the
fundamental gravitational scale will be higher, and 4D quantum
gravity effects will not emerge at the TeV scale. On the other
hand, ~\cite{CDE} considered lowering the
fundamental scale of the brane tension, and the 5D cosmological
constant to a TeV, with the fundamental Planck scale around $10^5$ TeV.
In this case, one can still use the warp factor
to generate the Planck scale. However, in this scenario, one has
very light KK modes, in fact of order $mm^{-1}$! So here, the
sub-millimeter phenomenology is similar to the large dimension
scenario.

We conclude that TeV experiments have the potential to unravel the
physics of extra dimensions, should we reach the higher
dimensional gravitational scale.

\section*{Acknowledgments}
We would like to thank Tony Gherghetta, Thomas Gregoire, Markus Luty, Riccardo
Rattazzi,  Valery Rubakov, Martin Schmaltz, Raman Sundrum, Jay
Wacker and Ed Witten for useful discussions. This work was supported in part by
the U.S. Department of Energy under contract DE-AC03-76SF00098 and
by the National Science Foundation under grant PHY-95-14797. NAH
is partially supported by the Alfred P. Sloan foundation and the David and 
Lucile Packard foundation, and thanks T. Gregoire and J. Wacker for 
discussion on the radion. MP is
supported in part by NSF grants PHY-97-22083 and PHY-00-070787. MP
 would like to thank LBL and CERN for their hospitality
during various phases of this work, and R. Rattazzi for useful discussons on 
his unpublished comments on radion stabilization.

\end{document}